\documentclass[fleqn,10pt]{wlscirep}
\usepackage[utf8]{inputenc}
\usepackage[T1]{fontenc}
\title{Fast barrier-free switching in synthetic antiferromagnets}

\author[1,2,3]{Yu.~Dzhezherya}
\author[1,2,3]{V.~Kalita}
\author[1,4]{P.~Polynchuk}
\author[1,5]{A.~Kravets}
\author[5]{V.~Korenivski}
\author[6]{S.~Kruchinin}
\author[7*]{S.~Bellucci}

\affil[1]{Institute of Magnetism, NASU and MESU, Kyiv, 03142, Ukraine}
\affil[2]{National Technical University of Ukraine “Igor Sikorsky Kyiv Polytechnic Institute”, Kyiv, 03056, Ukraine}
\affil[3]{Institute of Physics, NASU, Kyiv, 03028, Ukraine}
\affil[4]{Donetsk Institute for Physics and Engineering named after O.O.~Galkin, NASU, Kyiv, 03028, Ukraine.}
\affil[5]{Nanostructure Physics, Royal Institute of Technology, 10691 Stockholm, Sweden}
\affil[6]{Bogolyubov Institute for Theoretical Physics, NASU, Kyiv, 03143, Ukraine.}
\affil[7]{INFN-Laboratori Nazionali di Frascati, Via E. Fermi, 54, 00044, Frascati, Italy.}
\affil[*]{Correspondence and requests for materials should be addressed  S.B., bellucci@lnf.infn.it}

\begin{abstract}
We analytically solve the Landau-Lifshitz equations for the collective magnetization dynamics in a synthetic antiferromagnet (SAF) nanoparticle and uncover a regime of barrier-free switching under a short small-amplitude magnetic field pulse applied perpendicular to the SAF plane. We give examples of specific implementations for forming such low-power and ultra-fast switching pulses. For fully optical, resonant, barrier-free SAF switching we estimate the power per write operation to be $ \sim 100 $~pJ, 10-100 times smaller than for conventional quasi-static rotation, which should be attractive for memory applications.
\end{abstract}

\begin{document}

\flushbottom

\maketitle

\section*{Introduction}

Magnetic nanostructures is a field of magnetism under active development, motivated by their unique properties not found in the bulk as well as their broad technological applications. One such magnetic nanostructure system is a synthetic antiferromagnet (SAF). It consists of a multilayer, in which ferromagnetic (F) layers are coupled in antiparallel (AP) across non-magnetic (NM) spacer layers, thus resembling an antiferromagnet (AFM) system \cite{Duine:2018}. Similar to conventional atomic AFM's, the magnetization of a SAF is zero in the absence of an externally applied magnetic field. The interlayer AFM-type coupling in SAF's is easily controlled by choosing the materials and/or geometry of the multilayered element, in contrast to classical AFM's where the inter-sublattice exchange is fixed for a given crystal.  

Trilayer SAF's have the simplest design and consist of two AP-coupled F layers \cite{Leal:1998} either via an indirect exchange of RKKY type \cite{Gruenberg:2001} or a dipolar AP-coupling in the case of NM spacers and sub-micrometer elements. Due to the flux closure in the AP state, the magnetostatic fringing fields of a SAF element are nearly zero and so is the cross-talk in dense arrays of such elements. This makes SAF's widely used in magnetic recording read heads \cite{Parkin:2003,Zhu:1999} as well as magnetic random access memory (MRAM) \cite{Parkin:2003,Slaughter:2002,Apalkov:2016}, both as reference layers \cite{Leal:1998,Berg:1996} and free/storage layers \cite{Hayakawa:2006,Han:2007,Smith:2008,Yakata:2009,Lee:2011,Firastrau:2013}. Devices using SAF's show higher stability to thermal agitation and a broader dynamic range \cite{Han:2007,Yakata:2009}.

SAF-MRAM elements are made with uniaxial anisotropy such that each has two stable magnetization states for encoding a digital bit of information. Both stable states have the individual F layers mutually AP-aligned and their switching involves a transition (rotation or toggling) between the two AP states. The speed and efficiency of such memory cell depends on the speed and ease of the switching of the SAF element, which therefore are the key performance characteristics. Two types of SAF switching are often used: magnetic field switching and spin-transfer-torque (STT) switching. In field-MRAM, the switching is performed by generating an in-plane magnetic field pulse (or a pulse sequence) from currents in the word and bit lines located next to the SAF element in the MRAM array \cite{Engel:2005,Durlam:2007}. STT-MRAM is based on switching SAF memory elements by driving through them spin-polarized current pulses \cite{Rizzo:2013,Slaughter:2012,Hosomi:2005,Kawahara:2007}, which transfer spin between a magnetically fixed reference layer and the free/storage layer, switching it parallel or antiparallel to the reference layer. The low and high resistance states of the cell (P/AP states) are used to electrically read out the bit state. Recently, spin-orbit type STT switching has been demonstrated using the spin Hall effect (SHE) in a heavy-element metal layer with a strong spin-orbit coupling \cite{Lau:2016,Bi:2017,Shi:2017}, onto which a storage element is deposited and is magnetically switched by the interlayer spin-orbit torque. 

SAF's, in contrast to conventional AFM's, have the F layers coupled relatively weakly. This makes possible low-field, essentially barrier-free switching of small elliptical SAF elements, the mechanism of which is detailed below and is based on using perpendicular-to-the-plane field pulses of specific duration and amplitude, tuned to be in sync with the intrinsic dynamics of the SAF trilayer. Such field pulses can be generated by, e.g., circularly polarized laser pulses via the inverse Faraday effect, concentrated into the individual SAF using an optical antenna \cite{Kruglyak:2005,Kruglyak:2007,Kimel:2004,Kimel:2005,Kimel:2009}, with the direction of switching determined by the chirality of the laser pulse. The barrier-free SAF switching method discussed herein has significant advantages over the existing switching methods, which typically are done quasi-statically and require rather higher fields/driving currents.

In our previous studies, we showed that by using magnetic field pulses of specific shape one can achieve fast, inertia-free SAF switching \cite{Gorobets:2003,Konovalenko:2009,Dzhezherya:2012,Dzhezherya:2013,Cherepov:2010,Cherepov:2011,Koop:2013,Koop:2017}. In particular, we showed that the system's spin dynamics has collective modes of acoustic and optical types \cite{Konovalenko:2009}, that SAF can behave analogous to the Kapitza pendulum \cite{Dzhezherya:2012}, and that an oscillating in-plane field can be used to resonantly switch the system between the its stable AP states \cite{Cherepov:2011,Koop:2013}. Our most recent studies \cite{Kravets:2014,Kravets:2015,Kravets:2016} showed that an improved performance and additional functionality can be obtained by incorporating thermo-magnetic control of the SAF's interlayer coupling, which opens up a range of future spintronic device applications to be discussed in separate publications. 

An AFM system with the Dzyaloshinskii interaction \cite{Kimel:2009} subject to laser-induced magnetic field pulses showed switching followed by a relatively long equilibration process. The relevant fast-field AFM dynamics \cite{Satoh:2010,Galkin:2008} was studied in the Andreev-Baryakhtar framework \cite{Andreev:1980,Baryakhtar:1985}.
  
In this work, we theoretically investigate the regime of essentially barrier-free switching in nanoscale SAF elements using low-amplitude field pulses as well as possible experimental implementations of such switching. We show that under a coherent rotation of the two F moments in the AP state with zero total magnetization, the shape anisotropy of the elliptical particle has practically no effect of the spin dynamics of the system. This effectively eliminates the shape-induced magnetostatic barrier and makes switching fast and low power, which is desirable in various technological applications such as MRAM. 

\section*{Model formulation and results}

\begin{figure}[ht]
\centering
\includegraphics[width=6.5cm]{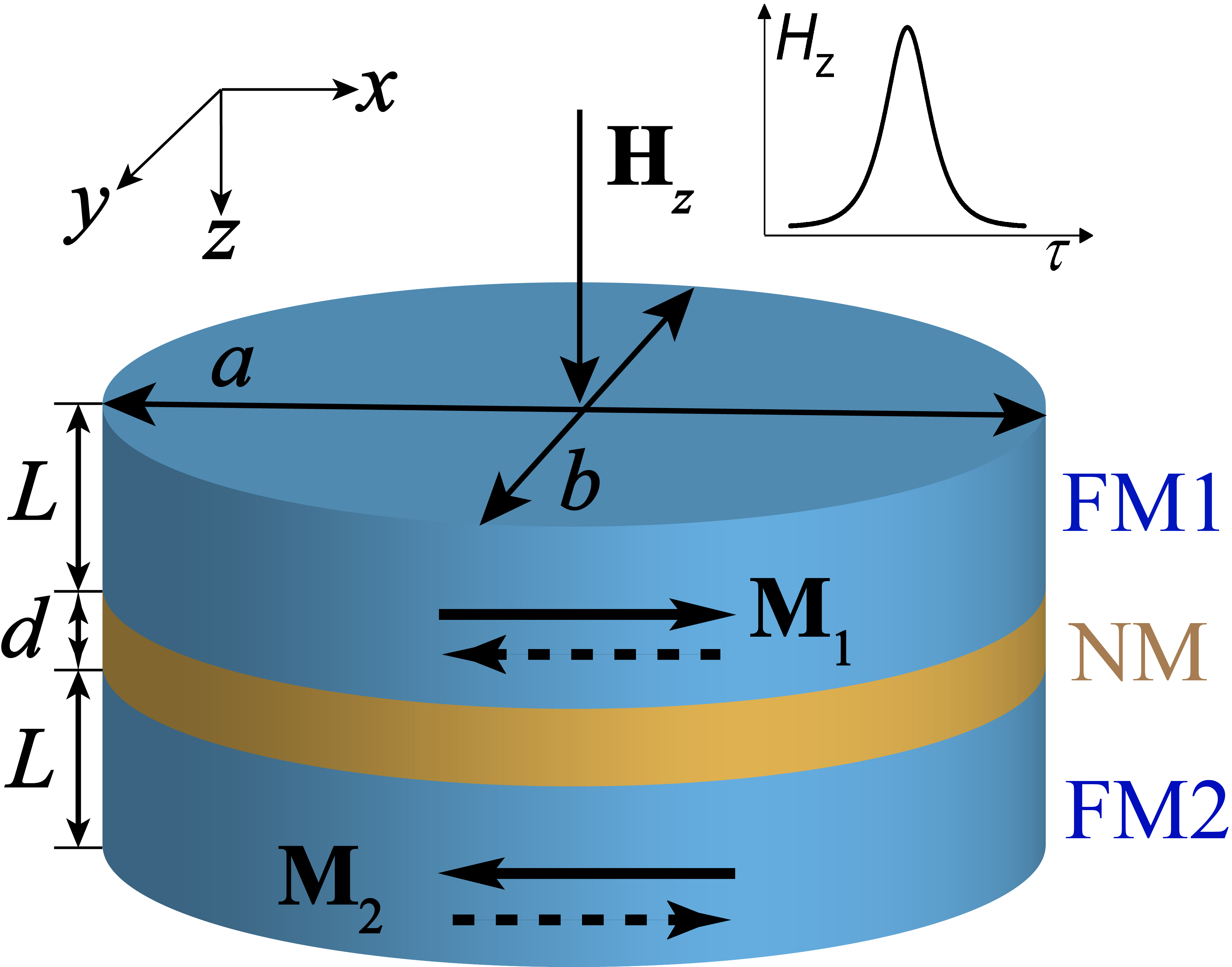}
\caption{Schematic of SAF cell. Two ferromagnetic layers with magnetizations \textbf{M$_1$} and \textbf{M$_2$} and thickness $L$ are separated by nonmagnetic spacer with thickness $d$. A pulse of out-of-plane excitating magnetic field \textbf{H$_z$} of certain duration and amplitude is produced either by external field coil or circularly polarized laser pulse (see Section: Switching  SAF optically).}
\label{Fig1}
\end{figure}

A SAF cell consists of two identical ferromagnetic thin-film elements (FM1 and FM2) separated by a nonmagnetic spacer layer (NM), as shown in Fig.~\ref{Fig1}. The magnetic moments \textbf{M$_1$} and \textbf{M$_2$} of layers FM1 and FM2 are oriented along the long elliptical axis of the SAF and are anti-parallel (AP) in the ground state due to their dipolar interaction thus forming an artificial (synthetic) antiferromagnetic system. One can distinguish the following SAF states: i)~$ l_x = m_{1x} - m_{2x}$, ii)~$ - l_x = - m_{1x} + m_{2x}$, which can then be used to encode binary information. Here \textbf{m}$_i$ = \textbf{M}$_i$/$M_{i\text{s}}$ -- the unit vector of the magnetization in $i$-th layer $\left( i = 1,\,2 \right)$ and $ M_{1\text{s}} = M_{2\text{s}} = M_\text{s}$ -- the saturation magnetization of the $i$-th layer. 

The task of writing information becomes that of developing efficient mechanisms of switching the SAF between the $ {l_x} $ and $ {-l_x}$  states. 

Here, we show how writing can be made most efficient by making the SAF elements rotate in the same direction during switching, retaining the AP orientation. Such rotation maintains the net magnetic moment of the cell at zero, which effectively cancels the magnetostatic barrier due the shape anisotropy.

The transverse SAF dimensions are assumed to be smaller than the characteristic magnetic length in the material, $ a \ll \Lambda =  \left( \sqrt{\alpha / 4 \pi } \right) $ ($\alpha$ -- exchange constant, $ a $ and $ b $ -- long and short half-axes of the elliptical SAF layers, see Fig.~\ref{Fig1} and therefore the magnetization in the SAF elements is considered to be uniform. We also assume that the magnetic layer thickness $L$ and eccentricity (aspect ratio close to 1) are sufficiently small. These conditions translate into the following applicability criterion for the developed theory:
\begin{equation}\label{eq:1}
   \Lambda \gg a \gg L, \qquad \left(a-b \right) / a \ll 1,
\end{equation}
where $a, b$ are the semi-axes of the SAF ellipse defined in Fig.~\ref{Fig1}. Criterion (\ref{eq:1}) allows to significantly lower the computational difficulties without restricting the general character of the results obtained. 

Taking into account the above considerations, the magnetic energy of the SAF system can be written as
\begin{equation} \label{eq:2}
   W = 4 \pi M_{\text{s}}^2 V \left\lbrace \sum \limits_{i=1}^2
   \left[ \frac{1}{2} \left[ N_{x}^{V_i} m_{ix}^2 + N_{y}^{V_i} m_{iy}^2 + \left( N_z^{V_i}-\beta_i \right) m_{iz}^2  \right]  - h{m_{iz}} \right] + A \textbf{m}_1\textbf{m}_2 + \sum \limits_{\alpha}\gamma_{\alpha} m_{1{\alpha}} m_{2{\alpha}} \right\rbrace,
\end{equation}
where $M_\text{s}$ -- saturation magnetization of the SAF layers; $V = V_1 = V_2 $ -- volumes of magnetic layers; $\bf{m}$$_i$ -- the unit vector of the magnetization in $i$-th layer $\left(i = 1, 2 \right)$; $\alpha = x, y, z$; $h = H_z/4\pi M_\text{s}$ -- normalized external magnetic field directed perpendicular to the SAF plane.

The parameter $4 \pi \beta_i$ determines the value of the uniaxial magnetic anisotropy of the $i$-th layer directed along the $O_z$ axis, since for thin layers the only selected direction is the normal to the surface. It should be noted that in the framework of this study, uniaxial anisotropy is not of fundamental importance, however, when forming SAFs with a layered structure, it can be induced due to the difference in the lattice periods of the contacting layers. To simplify further calculations, we will assume that the difference in the anisotropy of the magnetic layers is negligible and replace the values $\beta_i$ with some average value of $\beta$.

The term $A \textbf{m}_1\textbf{m}_2$, introduced into the expression for the energy density (\ref{eq:2}), describes the isotropic exchange interaction between the magnetic layers; it can differ in origin and magnitude. In the case of closely spaced layers, this interaction can be due to the diffusion of spin-polarized electrons \cite{Gorobets:2000,Kravets:2014} or the RKKY-type interaction \cite{Gruenberg:2001}. We note that by choosing the material and thickness of the non-magnetic spacer in such SAF structures one can influence the sign and magnitude of the effective interlayer coupling parameter, $A$.

Parameters $N_\alpha^{V_i}$ determine the demagnetization coefficients of the $i$-th layer, and $\gamma_\alpha$ -- the constants of the interlayer dipolar coupling ($\alpha =x,\, y,\, z$). The dipole-dipole interaction is always present in a SAF-particle system and will be used in this work to device a mechanism for controlling the SAF magnetization state. The values of the parameters of the dipole-dipole interaction are determined by the following formulas:
\begin{equation} \label{eq:3}
  N_{\alpha}^{V_i} = \frac{1}{4 \pi V} \left( {I_{V_i}^{V_i}} \right)_\alpha, 
    \qquad \gamma _{\alpha}  = \frac{1}{4\pi V} \left( I_{V_1}^{V_2} \right)_\alpha, 
    \qquad \left( I_{V_i}^{V_j} \right)_\alpha =\int \limits_{V_i} \int \limits_{V_j} dV dV' \frac{\partial ^2}{\partial x_\alpha \partial {x'}_\alpha }  \frac{1}{\left| \bf{r} - \bf{r'} \right|}\,,
\end{equation}
where $ i,j = 1,\,2 $; $V = V_1 = V_2$ -- volumes of the magnetic layers, over which the integration is performed. The top and bottom indexes in $ \left( I_{V_i}^{V_j} \right)_\alpha $ denote the integration regions in the first and second integrals in Eq. (\ref{eq:3}). Obviously, the demag-coefficients are symmetric with respect to top-to-bottom index exchange. If the two indexes coincide, $V_i=V_j$, then $\left(I_{V_i}^{V_j} \right)_\alpha$ is proportional to the average demag-coefficient of the given volume, $N_\alpha ^{V_i}$. Otherwise, $V_i \ne {V_j}$ and the interlayer coupling coefficient becomes $\left( I_{V_1}^{V_2} \right)_\alpha = 4 \pi V\gamma_\alpha$.

Due the the thin film geometry, the SAF cell has $N_z^{V_i}~\approx~1$. Therefore, in small applied fields, $\left| h \right| \ll 1$, the normal magnetization components are small, $\left| m_{iz} \right| \ll 1$. Also small are the in-plane components of the demagnetization tensor, $N_x^{V_i}$, $N_y^{V_i} \ll 1$ (their exact values will be given below).

Introducing variables $m_{iz}$ and $\varphi _i$, related to the magnetization vector components via 
\begin{equation}\label{eq:4}
   {{\bf{m}}_i} = \left( {\cos {\varphi _i}\sqrt {1 - m_{iz}^2} ,\,\,\,\,\sin {\varphi _i}\sqrt {1 - m_{iz}^2} ,\,\,\,\,{m_{iz}}} \right)
\end{equation}
and keeping the terms of not higher than second order of smallness, in analogy with \cite{Dzhezherya:2013}, the Lagrangian $\mathcal{L}$ in the form proposed \cite{Hubert:1974} becomes:
\begin{equation} \label{eq:5}
\begin{aligned}
  \mathcal{L}&= T - U,\\
    T& = -4\pi M_\text{s}^2V\left(\frac{2m_z}{\omega_M}\frac{d\Phi}{dt}+\frac{2l_z}{\omega_M} \frac{d\chi}{dt} \right),\\
       U&=4\pi M_\text{s}^2V \Bigl\{ \left(A+\overline{\gamma}\right)\cos2\chi -\frac{\cos2\Phi}{2} \left[ \gamma_y-\gamma_x+\left(N_y-N_x\right)\cos2\chi \right] +l_z^2 \left(1-\beta-2A\cos^2\chi\right) +m_z^2\left(1-\beta + 2A\sin^2\chi\right)-2m_z h \Bigr\},\\
          \mathcal{L}& = 4\pi M_\text{s}^2V \Biggl\{-\frac{2 m_z}{\omega_M} \frac{d \Phi}{dt} - \frac{2l_z}{\omega_M} \frac{d \chi}{dt} -\left (A + \overline{\gamma} \right) \cos2\chi +\frac{\cos2\Phi}{2}\left[\gamma_y-\gamma_x+\left(N_y-N_x \right)\cos2\chi\right] -l_z^2\left(1-\beta -2A \cos^2\chi \right)\\
   &\quad -m_z^2 \left( 1-\beta +2A \sin^2 \chi \right)+2m_zh \Biggr\}, \\
\end{aligned}
\end{equation}
where the following notations are introduced: $\omega_M = 8\pi\mu_0 M_\text{s}/\hbar$; $V=V_1=V_2$; $m_z =\left(m_{1z}+m_{2z}\right) / 2 $; $ l_z = \left(l_{1z} + l_{2z} \right) / 2 $; $ \Phi = \left( \varphi _1 + \varphi _2 \right) / 2 $; $ \chi = \left( \varphi _1 - \varphi _2 \right) / 2 $; $\overline{\gamma} = \left(\gamma_x + \gamma_y \right)/2$; $\mu_0$ is the Bohr magneton; it is also assumed that the demagnetization coefficients of both magnetic layers in the SAF are the same $ N_{\alpha}^{V_1} = N_{\alpha}^{V_2} = N_{\alpha}$.

Taking into account the damping in the Gilbert form can be done by introducing a dissipative function  
\begin{equation} \label{eq:6}
  \mathcal{R} = \frac{\alpha_\text{G} M_\text{s} V}{2g} \sum \limits_{i=1}^2 \left( \dot {\textbf{m}}_i\right)^2
\end{equation}
which in new variables is equivalent to the expression:
\begin{equation} \label{eq:7}
\begin{aligned}
  \mathcal{R} = 4\pi M_\text{s}^2V\frac{\alpha_\text{G}}{\omega_M}\Biggl\{\left(\frac{dm_z}{dt}\right)^2+\left({\frac{dl_z}{dt}}\right)^2 +\left(\frac{d\chi}{dt}\right)^2+\left(\frac{d\Phi}{dt}\right)^2\Biggr\},
\end{aligned}
\end{equation}
where $ \alpha _\text{G}$ is the dissipation constant in the Gilbert form. 
 
The four generalized coordinates, $q_\alpha$, of the Lagrange formalism in this case are $m_z$, $l_z$, $\Phi$ and $\chi$, and the equation system has the standard form:
\begin{equation} \label{eq:8}
  -\frac{d}{dt} \frac{\partial \mathcal{L}}{\partial {{\dot m}_z}} + \frac{\partial \mathcal{L}}{\partial m_z} = \frac{\partial \mathcal{R} }{\partial {\dot m_z}}, \quad  
  -\frac{d}{dt} \frac{\partial \mathcal{L}}{\partial {{\dot l}_z}}  + \frac{\partial \mathcal{L}}{\partial l_z} = \frac{\partial \mathcal{R} }{\partial {\dot l_z}}, \quad 
  -\frac{d}{dt} \frac{\partial \mathcal{L}}{\partial {{\dot \Phi}_z}} + \frac{\partial \mathcal{L}}{\partial \Phi_z} = \frac{\partial \mathcal{R} }{\partial {\dot \Phi_z}}, \quad 
  -\frac{d}{dt} \frac{\partial \mathcal{L}}{\partial {{\dot \chi}_z}} + \frac{\partial \mathcal{L}}{\partial \chi_z} = \frac{\partial \mathcal{R} }{\partial {\dot \chi_z}}.
\end{equation}
     
We separate the system of four equations (\ref{eq:8}) into two pairs:
\begin{subequations}\label{eq:9}
 \begin{eqnarray} 
 \frac{d \chi}{dt} + \left[ \omega_M \left( 1-\beta - 2A \cos^2\chi \right) + \alpha_\text{G} \frac{d}{dt} \right] l_z = & 0,
 \nonumber\\
 \frac{d l_z}{dt} - \alpha_\text{G} \frac{d \chi}{dt} + \omega_M \left[ A + \overline{\gamma} -\left(N_y-N_x \right)\cos2 \Phi \right] \sin\chi \cos \chi = & 0,
 \label{subeq:9a}
 \end{eqnarray}

  \begin{eqnarray}
  \frac{d\Phi}{dt}+\left[\omega_M\left(1-\beta + 2A\sin^2\chi\right]+\alpha_\text{G}\frac{d}{dt}\right]m_z = & \omega_Mh, 
  \nonumber\\
  -\frac{dm_z}{dt} + \alpha_\text{G}\frac{d\Phi}{dt}+\omega_M\left[\gamma_y-\gamma_x+\left(N_y-N_x \right)\cos2 \chi\right] \sin\Phi\cos\Phi = & 0.
  \label{subeq:9b}
  \end{eqnarray}
\end{subequations}

The system of two equations (\ref{subeq:9a}) is homogeneous -- contains no external excitation force. Moreover, for circular SAF's with $ N_x = N_{y}$, the systems of Eqs. (\ref{subeq:9a}) and (\ref{subeq:9b}) would become independent of one another. Then, Eqs. (\ref{subeq:9a}) would describe out-of-phase oscillations (so-called \textit{optical mode}) whereas Eqs. (\ref{subeq:9b}) would describe in-phase oscillations (\textit{acoustic mode}). Such basic SAF dynamics has been experimentally demonstrated and theoretically explained in detail in our earlier publications \cite{Korenivski:2005,Konovalenko:2009,Cherepov:2011,Dzhezherya:2012,Dzhezherya:2013}.

In this work, we show that dipolar SAF's with moderate in-plane shape anisotropy can exhibit a special type of switching, for which the potential barrier can be negligible. Moreover, under certain conditions, a parametric optical resonance can take place. 

The magnetostatic interaction in a planar nanosized SAF results in an antiferromagnetic ground state along the SAF's the long axis. A presence of additional interlayer exchange with $ A > 0 $ would lead to increased stability of the antiferromagnetic configuration. It is easy to show that the systems of Eqs.~(\ref{subeq:9a}) has a solution $l_z = 0$, $ x = \pm \pi/2 $, which corresponds to the antiferromagnetic state.

We will next show that a magnetic field pulse directed perpendicular to the SAF plane can switch the direction of the two magnetic moments while their mutual AF order is preserved during the switching process.

Let us consider the dynamic equations for the second pair of variables $ m_z $ and $ \Phi $ (Eq.~\ref{subeq:9b}) for the AP state, $\left( l_z=0,\,\,\chi= \pi/2 \right)$. Clearly, the system of two equations (\ref{subeq:9b}) is equivalent to one equation of second order that describes the synchronous rotation of the magnetization of the layers in one direction:
\begin{equation} \label{eq:10}
\begin{aligned}
  \frac{d^2 \Phi}{d \tau^2} &+ \alpha_\text{G} \left(1-\beta +2A \right) \left[ 1-\left( \frac{\Omega_s}{\left(1-\beta +2A \right)\omega_M}\right)^2 \cos2\Phi\right]\frac{d\Phi}{d\tau} - \left( \frac{\Omega_s}{\omega_M}\right)^2\sin\Phi\cos\Phi =  \frac{dh}{d\tau}, \\
  \frac{\Omega_s}{\omega_M} &= \sqrt {\left(1-\beta + 2A \right) \left( N_y-N_x -\gamma_y+\gamma_x \right)}, \qquad \tau = t\omega_M .
\end{aligned}
\end{equation}
Here in Eq.~(\ref{eq:10}), the terms proportional to $\alpha_\text{G}^2$ have been neglected. As a rule, materials with low intrinsic dissipation are selected for SAF. Otherwise, effective control of the magnetization state cannot be achieved. Since Eq.~(\ref{eq:10}) does not include parameter $A$, it becomes obvious that the isotropic exchange  interaction between the layers, of any origin, does not affect the dynamics of the synchronous  rotation of the two magnetic layers.

The dynamics of SAF in an out-of-plane magnetic field is largely determined by parameter $\Omega$, which can be interpreted as the eigen-frequency of small oscillations of \textit{acoustic} type in the vicinity of the equilibrium state. We proceed to find $\Omega$ in analytical form.
	
In-plane circular SAF elements would have the demagnetization coefficients described by the spheroid formula~\cite{Landau:1984}:
\begin{equation} \label{eq:11}
\begin{aligned}
  N_z^{V_1}=\frac{1}{1-\delta^2} \left(1-\frac{\delta\arccos\delta}{\sqrt{1-\delta^2}}\right), \qquad N_x^{V_1}=N_y^{V_1}=\frac{1}{2}\left(1-N_z^{V_1}\right),\\
\end{aligned}
\end{equation}
where $\delta=c/a <1$, $a=b=R$.

In the limiting case of a thin disk $\delta\ll 1$ the following expansion is valid:
\begin{equation}\label{eq:12}
  N_x^{V_1} = N_y^{V_1} = \frac{\pi L}{8R}\left(1-\frac{2}{\pi}\frac{L}{R}\right).
\end{equation}

The presence of eccentricity leads to inhomogeneity in the magnetostatic field distribution within the SAF layer. At the same time, the condition of Eq. (\ref{eq:1}) of small eccentricity makes this inhomogeneity small, such that the following approximate relations hold:

\begin{equation} \label{eq:13}
   N_x^{V_1} = N_x^{V_2} = \frac{\pi L}{8a'} \left( 1- \frac{2}{\pi} \frac{L}{R} \right), \qquad
   N_y^{V_1} = N_y^{V_2} = \frac{\pi L}{8b'} \left( 1- \frac{2}{\pi} \frac{L}{R} \right), \qquad
   N_z^{V_1} = N_z^{V_2} = 1-N_x^{V_1}-N_y^{V_1}.
\end{equation}
Here $ a' $ and $ b' $ are the effective half-axes of the magnetic layers, which are selected such that Eqs. (\ref{eq:4}) correspond to the average demagnetization coefficients, $ R= \left( a+b \right) /2 $. The values of the effective half-axes $ a' $ and $ b' $ can be expected to be close to the actual geometrical SAF parameters.

After finding the functional form of the demagnetization coefficients, Eq.~(\ref{eq:13}), the magnetostatic coupling constants can be found without cumbersome calculations. Indeed, based on Eqs.~(\ref{eq:3}), it is easy to show that the interlayer coupling constant can be expressed as
\begin{equation} \label{eq:14}
\begin{aligned}
  \gamma_\alpha &= \frac{1}{4\pi V} \left( I_{V_1}^{V_2} \right)_\alpha
   = \frac{1}{8\pi V} \left[ \left( I_{V_1 + V_2 + \Delta V}^{V_1 + V_2 + \Delta V} \right) _\alpha  -\left( I_{V_1 + \Delta V}^{V_1+\Delta V} \right)_\alpha - \left( I_{V_2 + \Delta V}^{V_2 +\Delta V} \right)_\alpha + \left( I_{\Delta V}^{\Delta V} \right)_\alpha \right] = \\ 
   &= \frac{1}{2 L} \Bigl\{ \left( 2L + d \right) N_\alpha^{V_1+V_2+\Delta V} -\left( L + d \right) \left( N_\alpha^{V_1 + \Delta V} + N_\alpha^{V_2+\Delta V} \right) + N_\alpha^{\Delta V} d \Bigl\},
 \end{aligned}
\end{equation}
where the upper index in the demag-coefficients indicates the relevant volume, for which it is computed. For example, notation $ V_1+V_2+\Delta V $ indicates that $ N_\alpha^{V_1+V_2+\Delta V} $ is the demag-coefficient of an effective particle with volume equal to the total volume of the two magnetic layers, $ V_1 $ and $ V_2 $, and the spacer, $ \Delta V $. The interlayer coupling coefficients $ \gamma_\alpha $  are thus expressed through the demag-coefficients of the respective combinations of the SAF elements.
Solving Eq.~(\ref{eq:14}) using Eq.~(\ref{eq:13}) yields the following results:
\begin{equation}\label{eq:15}
  \gamma_x=N_{x}^{V_1} - \frac{L\left( L+3d /2 \right)}{a'\left( a+b \right)}, \qquad 
  \gamma_y=N_{y}^{V_1} - \frac{L\left( L+3d /2 \right)}{b'\left( a+b \right)}, \qquad
  \gamma_{x} + \gamma_{y} + \gamma_{z} = 0.
\end{equation}

The validity condition for Eq.~(\ref{eq:15}) is that the SAF thickness is small compared to its lateral dimensions, ($ 2L + d \ll b $).
	
We have thus obtained that the interlayer coupling constants $\gamma_\alpha$ are different from the demag-coefficients by a quantity of the second order of smallness, $\sim \left( L /a \right)^2 \ll N_\alpha^{V_1}$, which is the model precision accepted at the outset. This has an interesting effect of a vanishing potential barrier for a synchronous rotation of the SAF layers in the AP state.
	
Indeed, it follows from Eqs.~(\ref{eq:13}) and (\ref{eq:15}) that the coefficients in Eq.~(\ref{eq:10}) determining the height of the switching barrier are proportional to 
\begin{equation}\label{eq:16}
  \left( \frac{\Omega _s}{\omega _M} \right)^2 = N_y^{V_1} - N_x^{V_1} - \gamma _y + \gamma _x = \left( 1-\beta +2A \right) \frac{a' - b'}{a + b} \frac{L \left( L + 3d /2 \right)}{a b} \approx \frac{ \left (1-\beta +2A \right) }{2} \frac{\left(a-b \right)}{a}\frac{L\left(L+3d/2\right)}{a^2}.
\end{equation}

The symmetry of the problem and the preservation of the AP state during magnetization rotations results in that the quantity in Eq.~(\ref{eq:16}), in the limit of Eq.~(\ref{eq:1}), is of the third order of smallness. Using the characteristic SAF-cell parameters yields $ \left( \Omega _s / \omega _M \right)^2 \sim 10^{-3} \div 10^{-4}$.
	
At the same time, for a rotation of the magnetic moment in an isolated single-layer, for which $\gamma _\alpha = 0$, the height of the barrier is of the second order of smallness: 
\begin{equation*}
  \left( \frac{\Omega'_s }{\omega _M} \right)^2 = \left(1-\beta \right)\left( N_y - N_x\right) = \frac{\pi \left(1-\beta \right)}{8}\frac{\left( a - b \right)}{a} \frac{L}{a} \sim 10^{-2} \div 10^{-3}.   
\end{equation*}

References \cite{Dzhezherya:2012,Dzhezherya:2013} considered some of specific cases of switching a SAF cell. Here we present the general approach to solving the problem of control of the SAF magnetization state using short pulses of magnetic field of arbitrary configuration.
	
The aim is that a short pulse of a magnetic field would transform a cell from state $\varphi _1= 0,\;\; \varphi _2 = \pi ,\;\; \Phi  = \left(\varphi _1 + \varphi _2 \right) / 2 = \pi /2 $ to state $ \varphi'_1 = \pi ,\;\; \varphi'_2 = 2 \pi ,\;\; \Phi' = \left( \varphi'_1 + \varphi'_2 \right) / 2 = 3 \pi /2 $.
	
This task and the nature of the SAF magnetization rotation allow to formulate the main condition on the parameters of the magnetic field pulse. 

It should be noted that in practice it is difficult to produce a spatially localized and short in duration field pulse of a high amplitude. We will therefore assume that $\left| h \right| \ll 1$, and that short in duration means, according to Eq.~(\ref{eq:10}),
\[ \left| \frac{dh}{d\tau } \right| \sim \frac{h_0}{T \omega _M} \gg \frac{1}{2} \left( \frac{ \Omega _s}{\omega _M} \right)^2 \sim 10^{-3} \div 10^{-4}, \]  
where $ h_0 $ and $ T $ -- characteristic values of the pulse amplitude and duration, respectively. The pulse has an arbitrary shape described by function $h\left(\tau-\tau_0\right)$, symmetric with respect to $\tau=\tau_0$. The main requirement is that the amplitude and duration must be selected such that the integral area of the pulse is equal to
\begin{equation} \label{eq:17}
  \int\limits_{\tau_0 -T}^{\tau_0+T} h \left(\tau-\tau_0 \right) d\tau \approx \int\limits_{-\infty}^{+\infty} h \left(\tau-\tau_0 \right) d\tau = \pi.
\end{equation}

The above requirements are not expected to produce any significant limitations on the properties or functioning of a SAF-based system in controlling the magnetization switching.
	
We rewrite Eq.~(\ref{eq:10}) in the following form:
\begin{equation}\label{eq:18}
  \frac{d}{d\tau}\left( \frac{d\Phi}{d\tau}-h \right) + \alpha_\text{G}\left(1-\beta +2A \right) \left( \frac{d\Phi}{d\tau} - h \right) - \left( \frac{\Omega _s}{\omega_M} \right)^2 \sin\Phi \cos\Phi = -\alpha_\text{G}  \left(1-\beta +2A \right)h.
\end{equation}
Expressing the angle variable as
\begin{equation} \label{eq:19}
  \Phi = \Phi _0 + \Phi _1, \qquad \Phi_0 = \frac{\pi}{2}+\int\limits_{-\infty}^{\tau} h\left(\tau-\tau_0\right)d\tau , 
\end{equation}
we expand Eq.~(\ref{eq:18}) into terms no higher than linear in $\Phi_1$:
\begin{multline} \label{eq:20}
   \frac{d^2\Phi_1}{d\tau^2}+\alpha_\text{G} \left(1-\beta + 2A \right)\frac{d\Phi_1}{d\tau}+\frac{1}{2}\left( \frac{\Omega_s}{\omega_M}\right)^2\sin2\Phi_1 = -\alpha_\text{G} \left(1-\beta +2A\right)h+\frac{1}{2}\left(\frac{\Omega_s}{\omega_M}\right)^2 \\ \times\left(2\sin2\Phi_1\cos^2\Phi_0+\sin2\Phi_0\cos2\Phi_1\right).
\end{multline}

The right hand side of Eq.~(\ref{eq:20}) collects the fast-varying terms and are non-zero only in the short period during the action of the field pulse, in the vicinity of $ \tau_0 $. Integrating Eq.~(\ref{eq:20}) in the immediate vicinity of $\tau_0$ and assuming that no excitation in the system prior to the pulse, we find the initial conditions and formulate the Cauchy problem for the system's excitations:
\begin{equation} \label{eq:21}
  \frac{d^2 \Phi_1}{d \tau^2}+\alpha_\text{G}  \left(1-\beta +2A \right) \frac{d \Phi_1}{d\tau} +\frac{1}{2} \left(\frac{\Omega_s}{\omega_M}\right)^2 \sin 2\Phi_1 = 0, \Phi_1 = 0,  \qquad \frac{d \Phi_1}{d \tau} = - \pi \alpha_\text{G}  \left(1-\beta +2A \right) \,\, \left|_{\tau =\tau_0}. \right.
\end{equation}

Eq.~(\ref{eq:21}) for $ \Phi_1^2 \ll 1 $ becomes linear and is easily solved. The correction in $ \Phi_1 $ thus introduced is equal to
\begin{equation*}
  \Phi_1 = - \frac{\pi \alpha_\text{G} \left(1-\beta +2A \right) \omega_M}{\Omega_s} e^{-\alpha_\text{G} \left(1-\beta +2A \right)\omega_M \left(t-t_0\right)} \sin \left[\Omega_s \left(t-t_0\right)\right].
\end{equation*}

\begin{figure}[ht]
\centering
\includegraphics[width=8.5 cm]{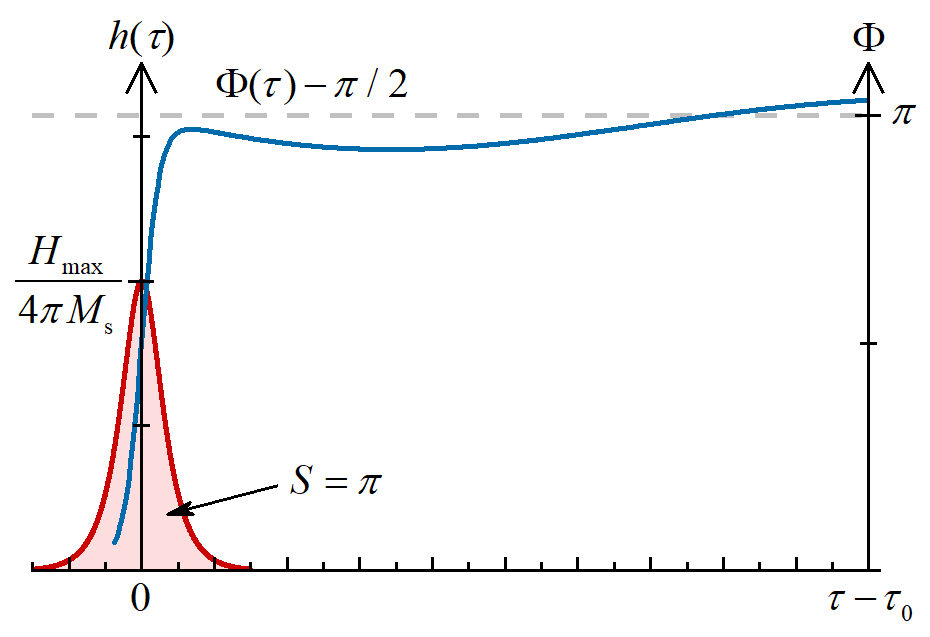}
\caption{Temporal profiles of external filed pulse and response of SAF cell representing switching between two stable AP states (SAF's antiferromagnetic vector angle changing from 0 to $ \pi $).}
\label{Fig2}
\end{figure}

Figure~\ref{Fig2} shows schematically the time profiles of the field pulse and the induced change of the SAF's antiferromagnetic vector angle, corresponding to switching between the two stable AP states of the cell.

Due to the approximation of $ \Phi_1^2 \ll 1 $ the validity condition for the developed theory becomes $ \bigl[ \pi \alpha _\text{G} \left(1-\beta +2A\right) \bigr]^2 \ll \left(\Omega_s / \omega_M \right)^2 $.

The resulting conditions for obtaining reliable pulse-induced switching of SAF take the following form:
\begin{subequations}\label{eq:22}
\begin{eqnarray}
  \int\limits_{-\infty}^{+\infty} h\left(\tau-\tau_0\right)d\tau = \pi, \label{subeq:22a} \\
  \frac{2\pi}{\left(\omega_M T \right)^2} \gg \frac{\left(1-\beta +2A \right)}{2}\frac{\left(a-b \right)}{a} \left(\frac{L}{a}\right)^2 \bigl[ \pi \alpha_\text{G} \left(1-\beta +2A \right) \bigr]^2. \label{subeq:22b}
\end{eqnarray}
\end{subequations}

Without presenting extensive calculations, for qualitative comparison, we show that a rotation of the magnetization vector of an isolated single-layer particle takes place at a much higher field amplitude $ h_0^\prime $. In this case, in expression $ \Omega_s / \omega_M =  \bigl[ \left( 1-\beta +2A \right)\left( N_y - N_x -\gamma_y +\gamma_x \right)  \bigr]^{1/2}$ the constants of the interlayer coupling are set to zero, $ \gamma_y$ = $ \gamma_x=0 $, and the switching condition takes the form:
\begin{subequations}\label{eq:23}
\begin{eqnarray}
  \int\limits_{-\infty}^{+\infty} h \left(\tau - \tau_0 \right)d \tau  = \pi, \label{subeq:23a} \\
  \frac{2\pi}{\left(\omega_M T\right)^2} \gg \frac{\pi L \left(1-\beta +2A \right)\left(a-b\right)}{8a^2} \gg  \bigl[\pi\alpha_\text{G} \left(1-\beta +2A \right) \bigr]^2. \label{subeq:23b}
\end{eqnarray}
\end{subequations}

From inequalities (\ref{subeq:22b}) and (\ref{subeq:23b}) it follows that a rotation of the two magnetic moments in a SAF such that their net in-plane magnetization is maintained at zero (AP, in-phase, \textit{acoustical} rotation) can be performed at a much lower field than a corresponding switching of a single-layer ferromagnetic particle. Given the same duration of the field pulse, the ratio of the amplitudes in the above two cases is equal to
\begin{equation} \label{eq:24}
  \frac{h_0}{h_0^\prime} = \frac{4}{\pi} \frac{L+3d/2}{a} \ll 1.
\end{equation}

This result is expected and is due to the potential barrier determined by the shape anisotropy being weakly expressed under a synchronous rotation of the two magnetic moments comprising the SAF, maintaining the AP state. 
	 
We note that fulfilling Eq.~(\ref{subeq:22a}) produces SAF switching in reverse, while Eq.~(\ref{subeq:22b}) is the criterion of high-speed and reliability of switching. In essence, Eq.~(\ref{subeq:22b}) simply ensures that the zeroth approximation deviates insignificantly from the exact time dependence of the magnetization rotation angle, $ \left| \Phi - \Phi_0 \right| \ll 1 $.
	
\begin{figure}[ht]
\centering
\includegraphics[width=14 cm]{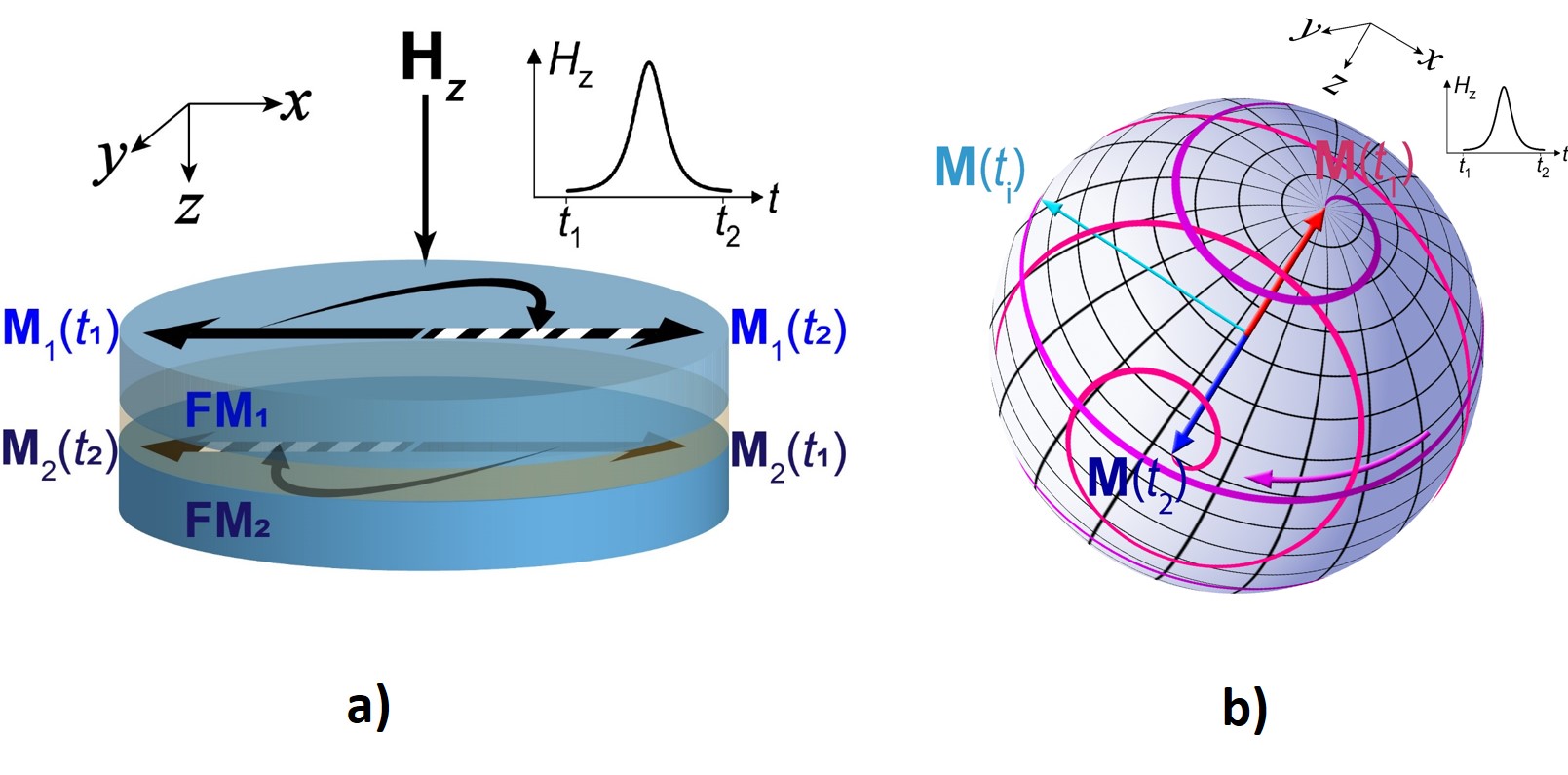}
\caption{Illustration of barrier-free and relaxation-free switching in SAF, coinciding in time with half-period of in-phase in-plane rotation (a), which is much faster than conventional out-of-phase switching of a ferromagnetic particle with subsequent precessional relaxation (b).}
\label{Fig3}
\end{figure}

Consequently, when Eq.~(\ref{eq:22}) is fulfilled, an out-plane field pulse produces a step-like switching of the SAF, with negligible post-switching relaxation (inertia-free switching). This SAF-resonant mechanism, analyzed in great detail in this work, is graphically illustrated in Fig.~\ref{Fig3}~a, and is qualitatively different and functionally superior compared to conventional switching of a ferromagnetic particle, illustrated in Fig.~\ref{Fig3}~b.
	
It should be informative to provide numerical estimates for the fast, low-amplitude pulses discussed above. We take the ferromagnetic material to be Nickel with the saturation magnetization $ M_\text{s} \sim 500 $~G and consequently $ \omega_M \approx 1.1 \cdot 10^{11}$~s$^{-1}$. We take the characteristic field pulse duration to be $T \sim 3 \cdot 10^{-10}$~s, such that $ \omega_M T \sim 33 $.

Condition (\ref{subeq:22b}) is valid for cells with parameters $ a/L \sim 10$, $\left( a - b \right)/a \sim 0.15 $, if the Gilbert damping constant $ \alpha _\text{G} < 10^{-2} $.

The condition of Eq.~(\ref{subeq:22a}) yields the amplitude of a field pulse needed for switching the SAF cell, given in Table~\ref{table_1} for several common pulse shapes. The results show that the amplitude does not depend on the saturation magnetization of the material and is determined by the shape and duration of the field pulse. 

\begin{table}[ht]
\centering
\begin{tabular} {|l|l|l|l|}
\hline
Case & Pulse shape, $ H_i(t)=4 \pi M_{\text{s}} h_i(t)$ & Amplitude, analytical $\left( H_i^{\max} \right)$ &  Amplitude, value at $T \sim 3 \times 10 ^{-10}$ s (Oe) \\
\hline
1 & $H_1^{\max}/\text{ch} (t/T))$ & $ \hbar/2\mu _0 T$ & 190 \\
\hline
2 & $H_2^{\max} / \text{ch}^2 (t/T)$ & $ \pi \hbar/4 \mu _0 T$ &  300 \\
\hline
3 & $H_3^{\max } /[1+(t/T)^2]$,  Lorentzian & $\hbar / 2\mu _0 T$ & 190 \\
\hline
4 & $H_4^{\max} \exp \left(-t/T\right)^2$ ,  Gaussian & $\sqrt \pi \hbar / 2 \mu _0 T$ & 330 \\
\hline
\end{tabular}
\caption{\label{table_1} Characteristic switching pulse amplitudes.}
\end{table}
 
\section*{Switching SAF optically}
 
Controlling the magnetization in nanoparticles using the magnetic field of a laser pulse was first considered in Ref.~\cite{Gorobets:2003}. The mechanism is based on the inverse Faraday effect first described in Ref.~\cite{Pitaevskii:1961}. This direction has since seen a broad spectrum of investigations~\cite{Kimel:2004,Kimel:2005,Kruglyak:2005,Kruglyak:2007,Kimel:2009}.
 
We previously showed \cite{Gorobets:2003} that the electric field of an electromagnetic wave having the following form in the SAF plane,
\begin{equation}\label{eq:25}
E_{x} = E_{1} \cos \omega_{E} t , \qquad E_{y} = E_{1} \cos \left( \omega_{E}t + \delta \right),
\end{equation}
(where $ \omega_E $ -- wave frequency, $ \delta $ -- phase shift between the oscillations along $ x,\,y $) leads to an effective circular motion of the electronic density in the conductive particle. Such electronic motion generates a magnetic moment, $ M_z^E $, and a magnetic field acting within the particle. At certain frequencies, depending on the shape of the particle, this effect can be amplified by the plasmon resonance of the conduction electrons.

The average value of the thus induced magnetization and magnetic field are given by the following expressions:
\begin{equation}\label{eq:26}
   \left \langle {M_z^E} \right \rangle = \sin \delta \frac{\omega E_1 E_2}{2cen_s} \left( \frac{1}{4 \pi N_\parallel} \right)^2  \left( \frac{\omega_s^2}{\omega_r^2 - \omega^2} \right)^2 \left(1+\frac{n_d}{n_s} \frac{m_s^2}{m_d^2} \right), \qquad 
  \left\langle H_z \right\rangle = - 4 \pi \left(1 - 2 N_\parallel \right)\left\langle M_z^E \right\rangle \approx - 4 \pi \left\langle M_z^E \right\rangle, 
\end{equation}
where $ e $ -- electron charge, $ n_l,\; m_l $ -- density and effective mass of the electrons in $ l $-th conduction zone, $\omega _l^2 = 4 \pi N_\parallel \left( e^2 n_l \right) / m_l $ -- characteristic frequency of the plasmonic oscillations in the $l$–th zone in the plane of the disk-shaped particle, $ N_\parallel \approx N_x^{V_1 + V_2 + \Delta V} \approx \pi \left( 2L+d \right)/ 8R $ -- average depolarization coefficient of the disk coinciding with the SAF's demagnetization coefficient, $ \omega_r = \left( \omega_s^2 + \omega_d^2 \right)^{1/2} \approx  \bigl[ 4 \pi N_\parallel e^2 \left( n_s / m_s + n_d / m_d \right)  \bigr]^{1/2} $ -- frequency of the plasmon resonance of the electron density in the SAF plane.

From the expression for $ \omega _r $ it follows that the effective means of regulating the plasmon resonance frequency in a SAF is tuning the ratio of the layers' thickness to the particle radius.  

For estimating the parameters of the pulse signal we take $ \omega _r \sim 3 \cdot 10^{15}$~s$^{-1} $, which is close to the ruby laser frequency. We further take $ \delta = \pi /2 $, and express the effective magnetic field of Eq.~(\ref{eq:26}) in the form
\begin{equation}\label{eq:27}
  \left\langle H_z \right\rangle =- \frac{I_0 \omega}{c^2 e n_s N_\parallel ^2} \left( \frac{\omega_s^2}{\omega_r^2 - \omega^2} \right)^2 \left( 1 + \frac{n_d}{n_s} \frac{m_s^2}{m_d^2} \right), \qquad
  I_0 =\frac{1}{2} \frac{c}{4 \pi } E^2.
\end{equation}

Assuming the time dependence of intensity $I_0$ corresponds to one of the cases presented in Table~\ref{table_1} and operation in the vicinity of the resonance frequency where $ \omega_s^2 / \left(\omega_r^2-\omega_E^2 \right)\sim 30 $, for $ n_s \sim 10^{22}$~cm$^{-3} $, $ N_\parallel \sim 0.1 $, $ \left\langle H_z \right\rangle\sim 200 $~Oe, we have the following estimate for the average density of the laser power flux needed to switch SAF:
\begin{equation} \label{eq:28}
  I_0 \sim 5\times10^{15} \text{erg} \, \text{c} ^{-1} \text{cm} ^{-2} = 5\times10^8 \, \text{W cm}^{-2}.
\end{equation}

The high, at first glance, value of the energy flux, when scaled by the pulse duration $ T \sim 3 \times 10^{-10} $~s, shows that the fraction of the light energy at the surface is only $ \sim 0.1 $ J~cm$^{-2} $. The area of a focused laser spot corresponding to that of a typical SAF cell is of the order of $ \Delta S \sim 10^{-9} $~cm$ ^{2} $, which translates to 100 pJ per write operation and is 10-100 times smaller compared to conventional over-the-barrier SAF switching.

Estimates show that in the case when a SAF grating is formed on a silicon substrate  of 1 cm$^2$ in area and 1 mm in thickness, for a single switching event per SAF element in the array of N~=~1~cm$^2$/$\lambda^2 \approx 2 \times 10^8$ elements, the temperature of the system would rise by an amount of the order of $\Delta T \approx 0.3$~K. 

These estimates indicate that, combined with the barrier-free regime of SAF-switching analyzed above, laser-induced normal-to-the-plane field pulsing can form the base of an efficient method of information writing on to SAF-type media

\section*{Appendix. Stability of magnetic states in SAF}

The stability of the magnetic states of SAF can be estimated from the value of the critical temperature, at which random thermal fluctuations of the magnetization are able to switch the antiferromagnetic SAF pair over the potential barrier and thus change the sign of the antiferromagnetic vector in the SAF plane.

We use the expression for the potential energy in the angular variables (\ref{eq:5}), assuming that the conditions of antiferromagnetic ordering $1 -\beta +2A > 0$, $ A + \overline{\gamma} >0$ are fulfilled and $\chi=\pi /2;\quad l_z=0$:

\begin{equation} \label{eq:29}
U=4\pi M_\text{s}^2V  \Bigl\{-\left( A + \overline{\gamma} \right) + \frac{\cos2\Phi}{2} \left(N_y - N_x -\gamma_y + \gamma_x \right) + m_z^2 \left( 1 - \beta +2A \right)-2 m_z h  \Bigl\}.
\end{equation}
\begin{figure}[ht]
\centering
\includegraphics[width=7 cm]{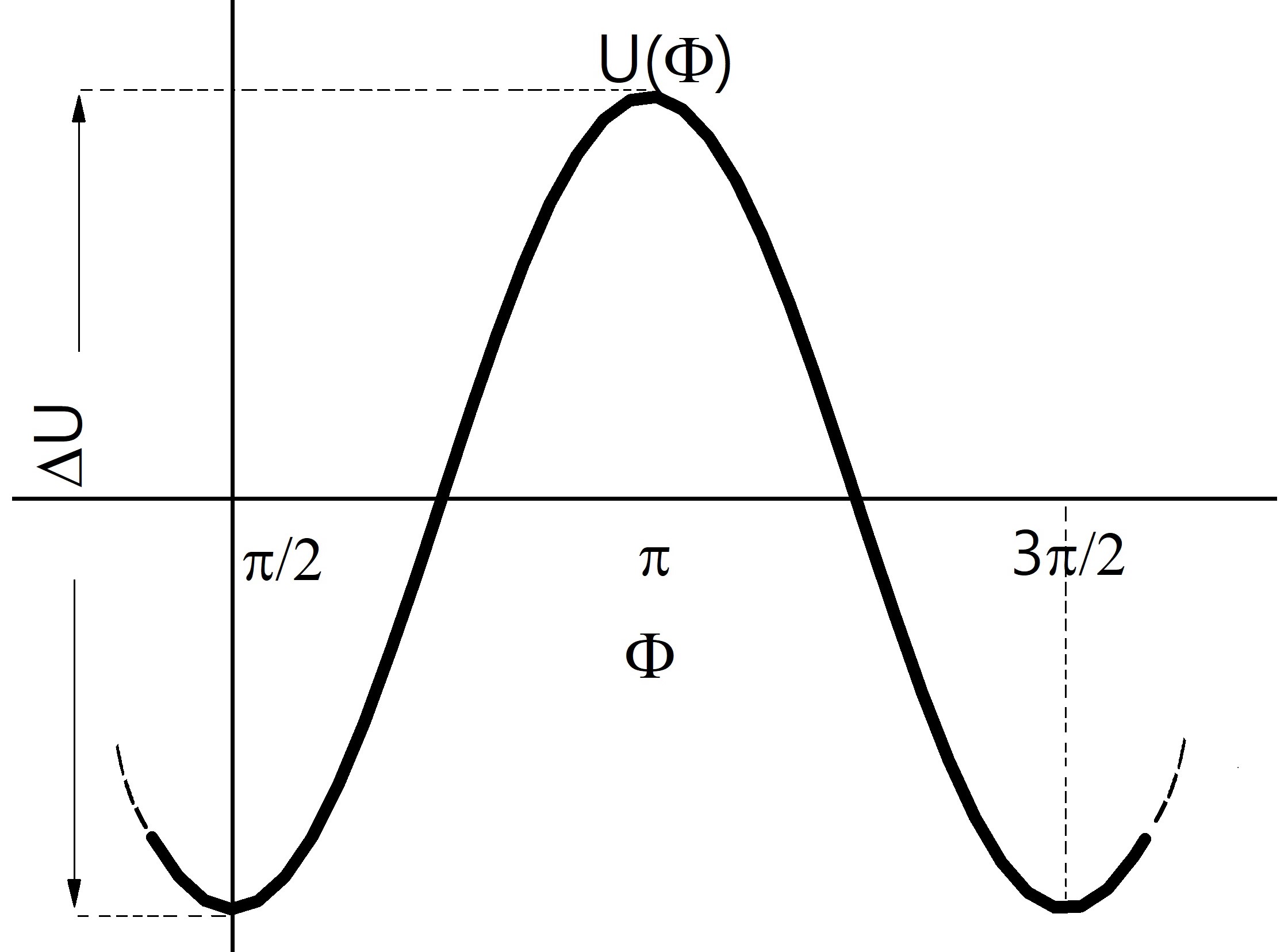}
\caption{Schematic of SAF's potential energy $U$ vs. angle $\Phi$.}
\label{Fig4}
\end{figure}

Figure~\ref{Fig4} shows schematically the dependence of the potential energy, $U$, on the angle, $ \Phi$.  As stated in the main text, angle $\Phi$ describes the synchronous (in phase) motion of the pair of magnetic moments, which is interpreted as oscillations of the ``acoustic type'' and can be considered as a ``soft mode'', since the potential barrier for the transition from one ground state, $\Phi_1 = \pi/2$, to the other, $\Phi_2=3\pi/2$, is relatively low:
\begin{equation}\label{eq:30}
\Delta U = 4 \pi M_\text{s}^2 V \left(N_y - N_x - \gamma_y + \gamma_x \right)= 4 \pi^2 M_\text{s}^2 L^2 \left(L + \frac{3d}{2} \right) \frac{a-b}{a+b}.
\end{equation}

One can estimate the critical temperature, $T_c$, at which thermal fluctuations are able to change the SAF's magnetic state: $T_\text{c} \approx \Delta U/k_\text{B}$ , where $k_\text{B}=1.38 \times 10^{-16}$ erg/K -- Boltzmann constant.

Taking the characteristic parameters of the SAF cell as $L = 5 \times 10^{-7}$~sm, $d=2 \times 10^{-7}$~sm, $a=3 \times 10^{-6}$~sm, $a-b=0.1a$, $M_\text{s} \approx 500$ G (Nickel), we obtain $T_\text{c} \approx 540$~K. This means that despite the relatively small potential barrier, the magnetic states of the SAF cell have high stability to thermal agitation. This fact is due to the large effective ``macrospin'' of the magnetic particles comprising the SAF.  When assessing the stability against thermal agitation, a nickel-based SAF ($M_\text{s}$) was considered as the low potential-barrier limit; widely used higher-$M_\text{s}$ ferromagnets such as Permalloy or CoFeB would yield higher thermal stability.

\bibliography{bibliography}

\begin{thebibliography}{10}
\urlstyle{rm}
\expandafter\ifx\csname url\endcsname\relax
  \def\url#1{\texttt{#1}}\fi
\expandafter\ifx\csname urlprefix\endcsname\relax\def\urlprefix{URL }\fi
\expandafter\ifx\csname doiprefix\endcsname\relax\def\doiprefix{DOI: }\fi
\providecommand{\bibinfo}[2]{#2}
\providecommand{\eprint}[2][]{\url{#2}}

\bibitem{Duine:2018}
\bibinfo{author}{Duine, R.~A.}, \bibinfo{author}{Lee, K.~J.}, \bibinfo{author}{Parkin, S. S.~P.} \& \bibinfo{author}{Stiles, M.~D.}
\newblock \bibinfo{journal}{\bibinfo{title}{Synthetic antiferromagnetic spintronics}}.
\newblock {\emph{\JournalTitle{Nat. Phys.}}} \textbf{\bibinfo{volume}{14}}, \bibinfo{pages}{217--219}, \doiprefix\url{10.1038/s41567-018-0050-y} (\bibinfo{year}{2018}).

\bibitem{Leal:1998}
\bibinfo{author}{Leal, J.~L.} \& \bibinfo{author}{Kryder, M.~H.}
\newblock \bibinfo{journal}{\bibinfo{title}{Spin valves exchange biased by co/ru/co synthetic antiferromagnets}}.
\newblock {\emph{\JournalTitle{J. Appl. Phys.}}} \textbf{\bibinfo{volume}{83}}, \bibinfo{pages}{3720--3723}, \doiprefix\url{10.1063/1.366597} (\bibinfo{year}{1998}).

\bibitem{Gruenberg:2001}
\bibinfo{author}{Gr\"{u}nberg, P.}
\newblock \bibinfo{journal}{\bibinfo{title}{Layered magnetic structures: History, highlights, applications}}.
\newblock {\emph{\JournalTitle{Phys. Today}}} \textbf{\bibinfo{volume}{54}}, \bibinfo{pages}{31--37}, \doiprefix\url{10.1063/1.1381100} (\bibinfo{year}{2001}).

\bibitem{Parkin:2003}
\bibinfo{author}{Parkin, S.} \emph{et~al.}
\newblock \bibinfo{journal}{\bibinfo{title}{Magnetically engineered spintronic sensors and memory}}.
\newblock {\emph{\JournalTitle{Proc. IEEE}}} \textbf{\bibinfo{volume}{91}}, \bibinfo{pages}{661--680}, \doiprefix\url{10.1109/jproc.2003.811807} (\bibinfo{year}{2003}).

\bibitem{Zhu:1999}
\bibinfo{author}{Zhu, J.~G.}
\newblock \bibinfo{journal}{\bibinfo{title}{Spin valve and dual spin valve heads with synthetic antiferromagnets}}.
\newblock {\emph{\JournalTitle{IEEE Trans. Magn.}}} \textbf{\bibinfo{volume}{35}}, \bibinfo{pages}{655--660}, \doiprefix\url{10.1109/20.750623} (\bibinfo{year}{1999}).

\bibitem{Slaughter:2002}
\bibinfo{author}{Slaughter, J.~M.} \emph{et~al.}
\newblock \bibinfo{journal}{\bibinfo{title}{Fundamentals of mram technology}}.
\newblock {\emph{\JournalTitle{J. Supercond.}}} \textbf{\bibinfo{volume}{15}}, \bibinfo{pages}{19--25}, \doiprefix\url{10.1023/a:1014018925270} (\bibinfo{year}{2002}).

\bibitem{Apalkov:2016}
\bibinfo{author}{Apalkov, D.}, \bibinfo{author}{Dieny, B.} \& \bibinfo{author}{Slaughter, J.~M.}
\newblock \bibinfo{journal}{\bibinfo{title}{Magnetoresistive random access memory}}.
\newblock {\emph{\JournalTitle{Proc. IEEE}}} \textbf{\bibinfo{volume}{104}}, \bibinfo{pages}{1796--1830}, \doiprefix\url{10.1109/jproc.2016.2590142} (\bibinfo{year}{2016}).

\bibitem{Berg:1996}
\bibinfo{author}{van~den Berg, H. A.~M.} \emph{et~al.}
\newblock \bibinfo{journal}{\bibinfo{title}{Gmr sensor scheme with artificial antiferromagnetic subsystem}}.
\newblock {\emph{\JournalTitle{IEEE Trans. Magn.}}} \textbf{\bibinfo{volume}{32}}, \bibinfo{pages}{4624--4626}, \doiprefix\url{10.1109/20.539099} (\bibinfo{year}{1996}).

\bibitem{Hayakawa:2006}
\bibinfo{author}{Hayakawa, J.} \emph{et~al.}
\newblock \bibinfo{journal}{\bibinfo{title}{Current-induced magnetization switching in mgo barrier based magnetic tunnel junctions with cofeb/ru/cofeb synthetic ferrimagnetic free layer}}.
\newblock {\emph{\JournalTitle{J. Appl. Phys.}}} \textbf{\bibinfo{volume}{45}}, \bibinfo{pages}{L1057--L1060}, \doiprefix\url{10.1143/jjap.45.l1057} (\bibinfo{year}{2006}).

\bibitem{Han:2007}
\bibinfo{author}{Han, J.~K.}, \bibinfo{author}{Shin, K.~H.} \& \bibinfo{author}{Lim, S.~H.}
\newblock \bibinfo{journal}{\bibinfo{title}{Thermal stability of a nanostructured trilayer synthetic antiferromagnet}}.
\newblock {\emph{\JournalTitle{J. Appl. Phys.}}} \textbf{\bibinfo{volume}{101}}, \bibinfo{pages}{09F506}, \doiprefix\url{10.1063/1.2710323} (\bibinfo{year}{2007}).

\bibitem{Smith:2008}
\bibinfo{author}{Smith, N.}, \bibinfo{author}{Maat, S.}, \bibinfo{author}{Carey, M.~J.} \& \bibinfo{author}{Childress, J.~R.}
\newblock \bibinfo{journal}{\bibinfo{title}{Coresonant enhancement of spin-torque critical currents in spin valves with a synthetic-ferrimagnet free layer}}.
\newblock {\emph{\JournalTitle{Phys. Rev. Lett.}}} \textbf{\bibinfo{volume}{101}}, \bibinfo{pages}{247205}, \doiprefix\url{10.1103/physrevlett.101.247205} (\bibinfo{year}{2008}).

\bibitem{Yakata:2009}
\bibinfo{author}{Yakata, S.} \emph{et~al.}
\newblock \bibinfo{journal}{\bibinfo{title}{Thermal stability and spin-transfer switchings in mgo-based magnetic tunnel junctions with ferromagnetically and antiferromagnetically coupled synthetic free layers}}.
\newblock {\emph{\JournalTitle{Appl. Phys. Lett.}}} \textbf{\bibinfo{volume}{95}}, \bibinfo{pages}{242504}, \doiprefix\url{10.1063/1.3275753} (\bibinfo{year}{2009}).

\bibitem{Lee:2011}
\bibinfo{author}{Lee, S.~W.} \& \bibinfo{author}{Lee, K.~J.}
\newblock \bibinfo{journal}{\bibinfo{title}{Current-induced magnetization switching of synthetic antiferromagnetic free layer in magnetic tunnel junctions}}.
\newblock {\emph{\JournalTitle{J. Appl. Phys.}}} \textbf{\bibinfo{volume}{109}}, \bibinfo{pages}{07C904}, \doiprefix\url{10.1063/1.3562214} (\bibinfo{year}{2011}).

\bibitem{Firastrau:2013}
\bibinfo{author}{Firastrau, I.}, \bibinfo{author}{Buda-Prejbeanu, L.~D.}, \bibinfo{author}{Dieny, B.} \& \bibinfo{author}{Ebels, U.}
\newblock \bibinfo{journal}{\bibinfo{title}{Spin-torque nano-oscillator based on a synthetic antiferromagnet free layer and perpendicular to plane polarizer}}.
\newblock {\emph{\JournalTitle{J. Appl. Phys.}}} \textbf{\bibinfo{volume}{113}}, \bibinfo{pages}{113908}, \doiprefix\url{10.1063/1.4795160} (\bibinfo{year}{2013}).

\bibitem{Engel:2005}
\bibinfo{author}{Engel, B.~N.} \emph{et~al.}
\newblock \bibinfo{journal}{\bibinfo{title}{A 4-mb toggle mram based on a novel bit and switching method}}.
\newblock {\emph{\JournalTitle{IEEE Trans. Magn.}}} \textbf{\bibinfo{volume}{41}}, \bibinfo{pages}{132--136}, \doiprefix\url{10.1109/tmag.2004.840847} (\bibinfo{year}{2005}).

\bibitem{Durlam:2007}
\bibinfo{author}{Durlam, M.} \emph{et~al.}
\newblock \bibinfo{title}{Toggle mram: A highly-reliable non-volatile memory}.
\newblock In \emph{\bibinfo{booktitle}{2007 International Symposium on VLSI Technology, Systems and Applications (VLSI-TSA)}}, \doiprefix\url{10.1109/vtsa.2007.378942} (\bibinfo{publisher}{IEEE}, \bibinfo{year}{2007}).

\bibitem{Rizzo:2013}
\bibinfo{author}{Rizzo, N.~D.} \emph{et~al.}
\newblock \bibinfo{journal}{\bibinfo{title}{A fully functional 64 mb ddr3 st-mram built on 90 nm cmos technology}}.
\newblock {\emph{\JournalTitle{IEEE Trans. Magn.}}} \textbf{\bibinfo{volume}{49}}, \bibinfo{pages}{4441--4446}, \doiprefix\url{10.1109/tmag.2013.2243133} (\bibinfo{year}{2013}).

\bibitem{Slaughter:2012}
\bibinfo{author}{Slaughter, J.~M.} \emph{et~al.}
\newblock \bibinfo{title}{High density st-mram technology (invited)}.
\newblock In \emph{\bibinfo{booktitle}{2012 International Electron Devices Meeting}}, \doiprefix\url{10.1109/iedm.2012.6479128} (\bibinfo{publisher}{IEEE}, \bibinfo{year}{2012}).

\bibitem{Hosomi:2005}
\bibinfo{author}{Hosomi, M.} \emph{et~al.}
\newblock \bibinfo{title}{A novel nonvolatile memory with spin torque transfer magnetization switching: spin-ram}.
\newblock In \emph{\bibinfo{booktitle}{IEEE International Electron Devices Meeting, 2005. IEDM Technical Digest.}}, \doiprefix\url{10.1109/iedm.2005.1609379} (\bibinfo{publisher}{IEEE}, \bibinfo{year}{2005}).

\bibitem{Kawahara:2007}
\bibinfo{author}{Kawahara, T.} \emph{et~al.}
\newblock \bibinfo{title}{2mb spin-transfer torque ram (spram) with bit-by-bit bidirectional current write and parallelizing-direction current read}.
\newblock In \emph{\bibinfo{booktitle}{2007 IEEE International Solid-State Circuits Conference. Digest of Technical Papers}}, \doiprefix\url{10.1109/isscc.2007.373503} (\bibinfo{publisher}{IEEE}, \bibinfo{year}{2007}).

\bibitem{Lau:2016}
\bibinfo{author}{Lau, Y.~C.}, \bibinfo{author}{Betto, D.}, \bibinfo{author}{Rode, K.}, \bibinfo{author}{Coey, J. M.~D.} \& \bibinfo{author}{Stamenov, P.}
\newblock \bibinfo{journal}{\bibinfo{title}{Spin-orbit torque switching without an external field using interlayer exchange coupling}}.
\newblock {\emph{\JournalTitle{Nat. Nanotech.}}} \textbf{\bibinfo{volume}{11}}, \bibinfo{pages}{758--762}, \doiprefix\url{10.1038/nnano.2016.84} (\bibinfo{year}{2016}).

\bibitem{Bi:2017}
\bibinfo{author}{Bi, C.} \emph{et~al.}
\newblock \bibinfo{journal}{\bibinfo{title}{Anomalous spin-orbit torque switching in synthetic antiferromagnets}}.
\newblock {\emph{\JournalTitle{Phys. Rev. B}}} \textbf{\bibinfo{volume}{95}}, \bibinfo{pages}{104434}, \doiprefix\url{10.1103/physrevb.95.104434} (\bibinfo{year}{2017}).

\bibitem{Shi:2017}
\bibinfo{author}{Shi, G.~Y.} \emph{et~al.}
\newblock \bibinfo{journal}{\bibinfo{title}{Spin-orbit torque in mgo/cofeb/ta/cofeb/mgo symmetric structure with interlayer antiferromagnetic coupling}}.
\newblock {\emph{\JournalTitle{Phys. Rev. B}}} \textbf{\bibinfo{volume}{95}}, \bibinfo{pages}{104435}, \doiprefix\url{10.1103/physrevb.95.104435} (\bibinfo{year}{2017}).

\bibitem{Kruglyak:2005}
\bibinfo{author}{Kruglyak, V.~V.} \& \bibinfo{author}{Portnoi, M.~E.}
\newblock \bibinfo{journal}{\bibinfo{title}{Generation of femtosecond current pulses using the inverse magneto-optical faraday effect}}.
\newblock {\emph{\JournalTitle{Tech. Phys. Lett.}}} \textbf{\bibinfo{volume}{31}}, \bibinfo{pages}{1047--1048}, \doiprefix\url{10.1134/1.2150894} (\bibinfo{year}{2005}).

\bibitem{Kruglyak:2007}
\bibinfo{author}{Kruglyak, V.~V.}, \bibinfo{author}{Portnoi, M.~E.} \& \bibinfo{author}{Hicken, R.~J.}
\newblock \bibinfo{journal}{\bibinfo{title}{Use of the faraday optical transformer for ultrafast magnetization reversal of nanomagnets}}.
\newblock {\emph{\JournalTitle{J. Nanophotonics}}} \textbf{\bibinfo{volume}{1}}, \bibinfo{pages}{013502}, \doiprefix\url{10.1117/1.2516174} (\bibinfo{year}{2007}).

\bibitem{Kimel:2004}
\bibinfo{author}{Kimel, A.~V.}, \bibinfo{author}{Kirilyuk, A.}, \bibinfo{author}{Tsvetkov, A.}, \bibinfo{author}{Pisarev, R.~V.} \& \bibinfo{author}{Rasing, T.}
\newblock \bibinfo{journal}{\bibinfo{title}{Laser-induced ultrafast spin reorientation in the antiferromagnet {TmFeO}3}}.
\newblock {\emph{\JournalTitle{Nature}}} \textbf{\bibinfo{volume}{429}}, \bibinfo{pages}{850--853}, \doiprefix\url{10.1038/nature02659} (\bibinfo{year}{2004}).

\bibitem{Kimel:2005}
\bibinfo{author}{Kimel, A.~V.} \emph{et~al.}
\newblock \bibinfo{journal}{\bibinfo{title}{Ultrafast non-thermal control of magnetization by instantaneous photomagnetic pulses}}.
\newblock {\emph{\JournalTitle{Nature}}} \textbf{\bibinfo{volume}{435}}, \bibinfo{pages}{655--657}, \doiprefix\url{10.1038/nature03564} (\bibinfo{year}{2005}).

\bibitem{Kimel:2009}
\bibinfo{author}{Kimel, A.~V.} \emph{et~al.}
\newblock \bibinfo{journal}{\bibinfo{title}{Inertia-driven spin switching in antiferromagnets}}.
\newblock {\emph{\JournalTitle{Nat. Phys.}}} \textbf{\bibinfo{volume}{5}}, \bibinfo{pages}{727--731}, \doiprefix\url{10.1038/nphys1369} (\bibinfo{year}{2009}).

\bibitem{Gorobets:2003}
\bibinfo{author}{Gorobets, Y.~I.}, \bibinfo{author}{Dzhezherya, Y.~I.} \& \bibinfo{author}{Kravets, A.~F.}
\newblock \bibinfo{journal}{\bibinfo{title}{Magnetization reversal of ferromagnetic granules by field of circularly-polarized electromagnetic waves}}.
\newblock {\emph{\JournalTitle{Metallofiz. Nov. Tekhn.}}} \textbf{\bibinfo{volume}{25}}, \bibinfo{pages}{27--36} (\bibinfo{year}{2003}).

\bibitem{Konovalenko:2009}
\bibinfo{author}{Konovalenko, A.}, \bibinfo{author}{Lindgren, E.}, \bibinfo{author}{Cherepov, S.~S.}, \bibinfo{author}{Korenivski, V.} \& \bibinfo{author}{Worledge, D.~C.}
\newblock \bibinfo{journal}{\bibinfo{title}{Spin dynamics of two-coupled nanomagnets in spin-flop tunnel junctions}}.
\newblock {\emph{\JournalTitle{Phys. Rev. B}}} \textbf{\bibinfo{volume}{80}}, \bibinfo{pages}{144425}, \doiprefix\url{10.1103/physrevb.80.144425} (\bibinfo{year}{2009}).

\bibitem{Dzhezherya:2012}
\bibinfo{author}{Dzhezherya, Y.~I.}, \bibinfo{author}{Demishev, K.~O.} \& \bibinfo{author}{Korenivskii, V.~N.}
\newblock \bibinfo{journal}{\bibinfo{title}{Kapitza problem for the magnetic moments of synthetic antiferromagnetic systems}}.
\newblock {\emph{\JournalTitle{JETP}}} \textbf{\bibinfo{volume}{115}}, \bibinfo{pages}{284--288}, \doiprefix\url{10.1134/s1063776112070047} (\bibinfo{year}{2012}).

\bibitem{Dzhezherya:2013}
\bibinfo{author}{Dzhezherya, Y.~I.}, \bibinfo{author}{Yurchuk, V.~P.}, \bibinfo{author}{Demishev, K.~O.} \& \bibinfo{author}{Korenivskii, V.~N.}
\newblock \bibinfo{journal}{\bibinfo{title}{Remagnetization of synthetic antiferromagnetic cells by a magnetic field pulse}}.
\newblock {\emph{\JournalTitle{JETP}}} \textbf{\bibinfo{volume}{117}}, \bibinfo{pages}{1059--1065}, \doiprefix\url{10.1134/s1063776113140100} (\bibinfo{year}{2013}).

\bibitem{Cherepov:2010}
\bibinfo{author}{Cherepov, S.~S.}, \bibinfo{author}{Korenivski, V.} \& \bibinfo{author}{Worledge, D.~C.}
\newblock \bibinfo{journal}{\bibinfo{title}{Resonant switching of two dipole-coupled nanomagnets}}.
\newblock {\emph{\JournalTitle{IEEE Trans. Magn.}}} \textbf{\bibinfo{volume}{46}}, \bibinfo{pages}{2112--2115}, \doiprefix\url{10.1109/tmag.2010.2043715} (\bibinfo{year}{2010}).

\bibitem{Cherepov:2011}
\bibinfo{author}{Cherepov, S.~S.}, \bibinfo{author}{Koop, B.~C.}, \bibinfo{author}{Dzhezherya, Y.~I.}, \bibinfo{author}{Worledge, D.~C.} \& \bibinfo{author}{Korenivski, V.}
\newblock \bibinfo{journal}{\bibinfo{title}{Resonant activation of a synthetic antiferromagnet}}.
\newblock {\emph{\JournalTitle{Phys. Rev. Lett.}}} \textbf{\bibinfo{volume}{107}}, \bibinfo{pages}{077202}, \doiprefix\url{10.1103/physrevlett.107.077202} (\bibinfo{year}{2011}).

\bibitem{Koop:2013}
\bibinfo{author}{Koop, B.~C.} \emph{et~al.}
\newblock \bibinfo{journal}{\bibinfo{title}{Demonstration of bi-directional microwave-assisted magnetic reversal in synthetic ferrimagnets}}.
\newblock {\emph{\JournalTitle{Appl. Phys. Lett.}}} \textbf{\bibinfo{volume}{103}}, \bibinfo{pages}{142408}, \doiprefix\url{10.1063/1.4824016} (\bibinfo{year}{2013}).

\bibitem{Koop:2017}
\bibinfo{author}{Koop, B.~C.}, \bibinfo{author}{Descamps, T.}, \bibinfo{author}{Holmgren, E.} \& \bibinfo{author}{Korenivski, V.}
\newblock \bibinfo{journal}{\bibinfo{title}{Relaxation-free and inertial switching in synthetic antiferromagnets subject to super-resonant excitation}}.
\newblock {\emph{\JournalTitle{IEEE Trans. Magn.}}} \textbf{\bibinfo{volume}{53}}, \bibinfo{pages}{1--5}, \doiprefix\url{10.1109/tmag.2017.2707589} (\bibinfo{year}{2017}).

\bibitem{Kravets:2014}
\bibinfo{author}{Kravets, A.~F.} \emph{et~al.}
\newblock \bibinfo{journal}{\bibinfo{title}{Synthetic ferrimagnets with thermomagnetic switching}}.
\newblock {\emph{\JournalTitle{Phys. Rev. B}}} \textbf{\bibinfo{volume}{90}}, \bibinfo{pages}{104427}, \doiprefix\url{10.1103/physrevb.90.104427} (\bibinfo{year}{2014}).

\bibitem{Kravets:2015}
\bibinfo{author}{Kravets, A.~F.} \emph{et~al.}
\newblock \bibinfo{journal}{\bibinfo{title}{Spin dynamics in a curie-switch}}.
\newblock {\emph{\JournalTitle{J. Phys. Condens. Matter}}} \textbf{\bibinfo{volume}{27}}, \bibinfo{pages}{446003}, \doiprefix\url{10.1088/0953-8984/27/44/446003} (\bibinfo{year}{2015}).

\bibitem{Kravets:2016}
\bibinfo{author}{Kravets, A.~F.} \emph{et~al.}
\newblock \bibinfo{journal}{\bibinfo{title}{Anisotropic magnetization relaxation in ferromagnetic multilayers with variable interlayer exchange coupling}}.
\newblock {\emph{\JournalTitle{Phys. Rev. B}}} \textbf{\bibinfo{volume}{94}}, \bibinfo{pages}{064429}, \doiprefix\url{10.1103/physrevb.94.064429} (\bibinfo{year}{2016}).

\bibitem{Satoh:2010}
\bibinfo{author}{Satoh, T.} \emph{et~al.}
\newblock \bibinfo{journal}{\bibinfo{title}{Spin oscillations in antiferromagnetic nio triggered by circularly polarized light}}.
\newblock {\emph{\JournalTitle{Phys. Rev. Lett.}}} \textbf{\bibinfo{volume}{105}}, \bibinfo{pages}{077402}, \doiprefix\url{10.1103/physrevlett.105.077402} (\bibinfo{year}{2010}).

\bibitem{Galkin:2008}
\bibinfo{author}{Galkin, A.~Y.} \& \bibinfo{author}{Ivanov, B.~A.}
\newblock \bibinfo{journal}{\bibinfo{title}{Dynamics of antiferromagnets exposed to ultrashort magnetic field pulses}}.
\newblock {\emph{\JournalTitle{JETP Lett.}}} \textbf{\bibinfo{volume}{88}}, \bibinfo{pages}{249--253}, \doiprefix\url{10.1134/s0021364008160054} (\bibinfo{year}{2008}).

\bibitem{Andreev:1980}
\bibinfo{author}{Andreev, A.~F.} \& \bibinfo{author}{Marchenko, V.~I.}
\newblock \bibinfo{journal}{\bibinfo{title}{Symmetry and the macroscopic dynamics of magnetic materials}}.
\newblock {\emph{\JournalTitle{Sov. Phys. Usp.}}} \textbf{\bibinfo{volume}{23}}, \bibinfo{pages}{21--34}, \doiprefix\url{10.1070/pu1980v023n01abeh004859} (\bibinfo{year}{1980}).

\bibitem{Baryakhtar:1985}
\bibinfo{author}{Baryakhtar, V.~G.}, \bibinfo{author}{Ivanov, B.~A.} \& \bibinfo{author}{Chetkin, M.~V.}
\newblock \bibinfo{journal}{\bibinfo{title}{Dynamics of domain walls in weak ferromagnets}}.
\newblock {\emph{\JournalTitle{Sov. Phys. Usp.}}} \textbf{\bibinfo{volume}{28}}, \bibinfo{pages}{563--588}, \doiprefix\url{10.1070/pu1985v028n07abeh003871} (\bibinfo{year}{1985}).

\bibitem{Gorobets:2000}
\bibinfo{author}{Gorobets, Y.~I.}, \bibinfo{author}{Dzhezherya, Y.~I.} \& \bibinfo{author}{Kravets, A.~F.}
\newblock \bibinfo{journal}{\bibinfo{title}{Magnetic ordering in granular system}}.
\newblock {\emph{\JournalTitle{Phys. Solid State}}} \textbf{\bibinfo{volume}{42}}, \bibinfo{pages}{126--131}, \doiprefix\url{10.1134/1.1131179} (\bibinfo{year}{2000}).

\bibitem{Hubert:1974}
\bibinfo{author}{Hubert, A.}
\newblock \emph{\bibinfo{title}{Theorie der Dom\"{a}nenw\"{a}nde in geordneten Medien}} (\bibinfo{publisher}{Springer Berlin, Heidelberg}, \bibinfo{year}{1974}).

\bibitem{Korenivski:2005}
\bibinfo{author}{Korenivski, V.} \& \bibinfo{author}{Worledge, D.~C.}
\newblock \bibinfo{journal}{\bibinfo{title}{Thermally activated switching in spin-flop tunnel junctions}}.
\newblock {\emph{\JournalTitle{Appl. Phys. Lett.}}} \textbf{\bibinfo{volume}{86}}, \bibinfo{pages}{252506}, \doiprefix\url{10.1063/1.1947907} (\bibinfo{year}{2005}).

\bibitem{Landau:1984}
\bibinfo{author}{Landau, L.~D.} \& \bibinfo{author}{Lifshitz, E.~M.}
\newblock \emph{\bibinfo{title}{Electrodynamics of continuous media}} (\bibinfo{publisher}{Pergamon, Oxford}, \bibinfo{year}{1984}).

\bibitem{Pitaevskii:1961}
\bibinfo{author}{Pitaevskii, L.~P.}
\newblock \bibinfo{journal}{\bibinfo{title}{Electric forces in a transparent dispersive medium}}.
\newblock {\emph{\JournalTitle{Sov. Phys. JETP}}} \textbf{\bibinfo{volume}{12}}, \bibinfo{pages}{1008--1013} (\bibinfo{year}{1961}).

\end{thebibliography}

\section*{Acknowledgements}

Support from the National Academy of Sciences of Ukraine (project 0121U108844), the Ministry of Education and Science of Ukraine (Project 0122U002233), the European Project H2020-MSCA-RISE-2017-778308 -- SPINMULTIFILM, the Swedish Research Council (VR 2018-03526), the Olle Engkvist Foundation (project 2020-207-0460), the Wenner-Gren Foundation (grant GFU2022-0011), and the Swedish Strategic Research Council (SSF UKR22-0050) are gratefully acknowledged.

\section*{Author contributions statement}
All authors contributed equally to the calculations in the work and to the writing of the article. Yu. Dzhezherya and S. Bellucci jointly supervised the work. All authors reviewed the manuscript.

\section*{Additional information}
\textbf{Competing interests}: The authors declare no competing interests. 

\end{document}